\documentclass[%
 aip,
 amsmath,amssymb,
 reprint,%
]{revtex4-1}

\usepackage{graphicx}
\usepackage{dcolumn}
\usepackage{bm}

\usepackage[utf8]{inputenc}
\usepackage[T1]{fontenc}
\usepackage{mathptmx}
\usepackage{etoolbox}
\usepackage{xcolor}

\newcommand{\bb}[0]{\begin{eqnarray}}
\newcommand{\ee}[0]{\end{eqnarray}}

\makeatletter
\def\@email#1#2{%
 \endgroup
 \patchcmd{\titleblock@produce}
  {\frontmatter@RRAPformat}
  {\frontmatter@RRAPformat{\produce@RRAP{*#1\href{mailto:#2}{#2}}}\frontmatter@RRAPformat}
  {}{}
}%
\makeatother
\begin{document}

\preprint{AIP/123-QED}

\title{Non-linear regime for enhanced performance of an Aharonov-Bohm heat engine}
\author{G\'eraldine Haack}
\affiliation{ 
Department of Applied Physics, University of Geneva, 1211 Gen\`eve, Switzerland
}
\author{Francesco Giazotto}%
 \email{geraldine.haack@unige.ch, francesco.giazotto@sns.it}
\affiliation{ 
NEST, Instituto Nanoscienze-CNR and Scuola Normale Superiore, I-56127 Pisa, Italy
}%

\date{\today}

\begin{abstract}
Thermal transport and quantum thermodynamics at the nanoscale is nowadays garnering an increasing attention, in particular in the context of quantum technologies. Experiments relevant for quantum technology are expected to be performed in the non-linear regime. In this work, we build on previous results derived in the linear response regime for the performance of an Aharonov-Bohm (AB) interferometer operated as heat engine. In the non-linear regime, we demonstrate the tunability, large efficiency and thermopower that this mesoscopic quantum machine can achieve, confirming the exciting perspectives that this AB ring offers for developing efficient thermal machines in the fully quantum regime.  
\end{abstract}

\maketitle

\section{Introduction}

Thermoelectricity in quantum system is nowadays gaining an increasing interest since the enormous advances achieved in quantum technology. In this context, a crucial question is related to determine if the performances of thermoelectric heat engines are affected by quantum mechanics, in particular, by influencing their output power and thermodynamic conversion efficiency, especially in the non-linear response regime \cite{Houten92, Giazotto06, Benenti17, Whitney18, Whitney14, Whitney15,  Lopez16}. Yet, it is well established that phase coherence and quantum effects do play a stark role  in governing the overall behavior of mesoscopic heat engines \cite{Blanter97, Hofer15, Lambert16, Samuelsson17, Hwang18, Haack19}. An emblematic example of phase-tunable quantum device is represented by the celebrated Aharonov-Bohm (AB) interferometer \cite{Aharonov59, Buttiker84, Yeyati95}, where the charge particle response can be influenced by either electric or magnetic potentials. AB-type interferometers represent suitable building blocks for the implementation of efficient and versatile phase-coherent quantum heat engines. In particular, in a previous work Ref.~[\onlinecite{Haack19}], the authors investigated the performance of AB-based quantum heat engines operated in the linear-response regime, highlighting the excellent tunability and behavior provided by magnetic fields and gate voltage control. However, experiments relevant for quantum technology applications often require the operation of a device well beyond the linear-response regime.

\begin{figure}[t]
 \centering 
\includegraphics[width=\columnwidth]{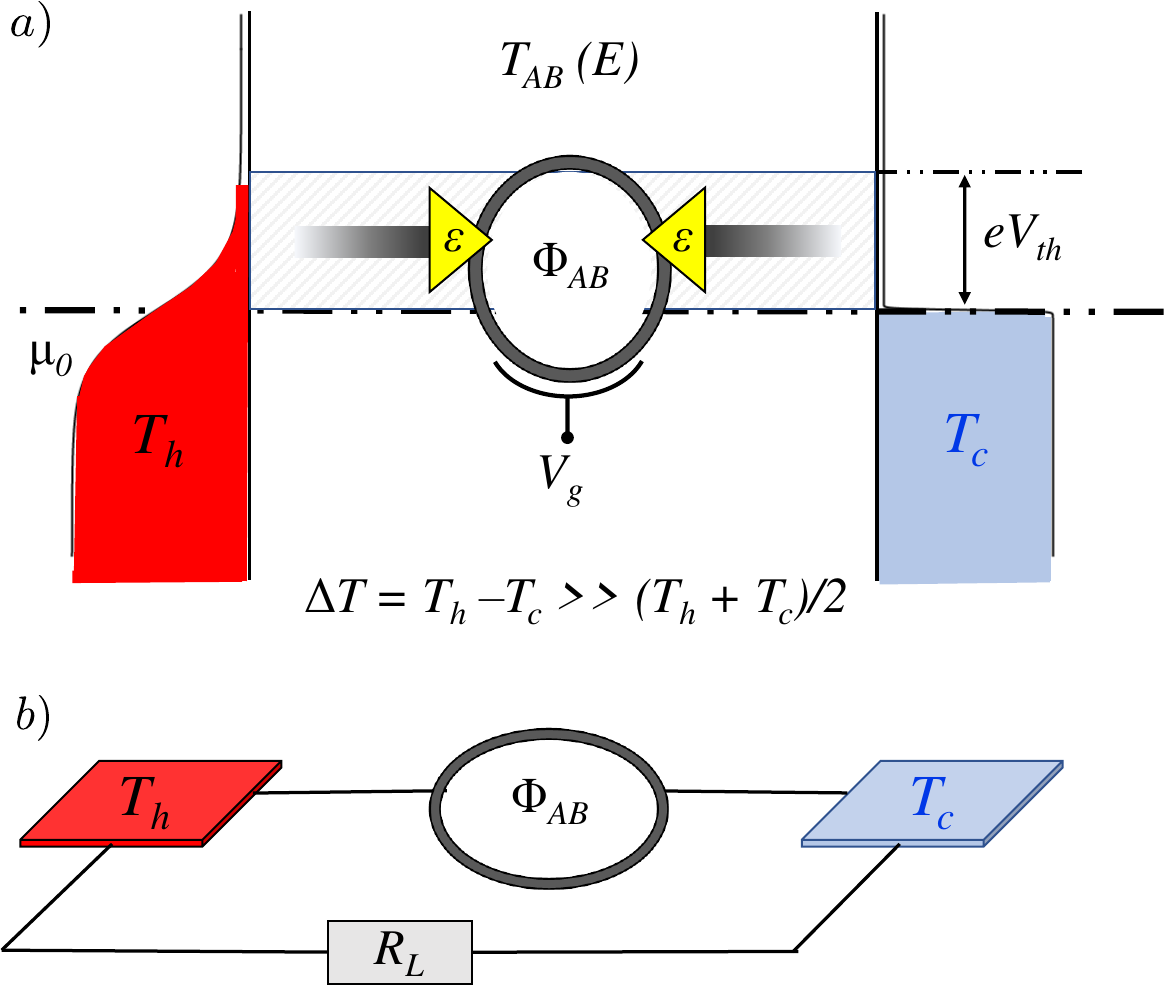}
\caption{\textbf{Panel a)}. Open-circuit configuration of the AB ring in the non-linear response regime. Left and right contacts are biased in temperatures, respectively at $T_{h}$ and $T_{c}$, and set to same chemical potential $\mu_0$. The tranmission probability of the AB ring $T_{AB}(E)$ is energy-dependent, and also depends on the length (time) imbalance between the two arms, on $V_g$, an external gate voltage, on the transmission parameter $\epsilon$ of the T-junctions and on $\Phi_{AB}$, the magnetic flux piercing the interferometer. Due to thermoelectric properties of the AB ring, a thermovoltage $V_{th}$ ($e V_{th}$ being the corresponding energy with $e$ the electrical charge) is developped between the contacts, in response to the temperature gradient.  \textbf{Panel b)}. Closed-circuit configuration for operating the AB ring as a quantum heat engine, when the interferometer is connected to a generic load resistor $R_L$ through dissipationless superconducting lines.}
\label{fig:model}
\end{figure}

Here, we build upon previous results derived in the linear-response regime by investigating the behavior of an AB interferometer operated in the full \textit{non-linear} regime as a quantum heat engine. We will demonstrate the good tunability intrinsic to the device, the sizable thermodynamic conversion efficiency, and the very large thermopower that this phase-controllable thermoelectric quantum machine can achieve. All this confirms the AB interferometer as a prototypical tool to develop efficient thermoelectric machines operating in the fully quantum regime.

After briefly recalling the model in Sec.~\ref{sec:model}, this contribution is organized as follows. In Sec.~\ref{sec:open}, we estimate within a scattering-matrix approach the thermoelectric properties of the AB ring in the non-linear response regime, through the thermovoltage and differential Seebeck coefficient. This is done in an open-circuit configuration, as a function of the AB magnetic flux, external gate voltage and temperature bias. We also discuss the relevance of our analysis with respect to the possible role of interactions in this device. In Sec.~\ref{sec:closed}, we turn to the closed-circuit configuration that allows us to operate the AB ring setup as a heat engine. Specifically, we provide pedagogical steps towards the characterization of a quantum thermal machine in the non-linear response regime. In Sec.~\ref{sec:eff}, we compute the heat current, power, efficiency and determine the optimal load resistance for a given temperature bias in order to maximize both power and efficiency of the AB heat engine. We conclude with perspectives towards efficient phase-coherent quantum thermal machines.

\section{Model}
\label{sec:model}

We start by introducing our model for the AB ring shown in Fig.~\ref{fig:model}. The AB ring is connected to two metallic contacts that can be biased in voltage or in temperature. Transmission probabilities between the contacts and the ring can be tuned for both contacts independently by controlling the transmission probability of the T-junctions \cite{Buttiker84}. Electrons traveling through the ring acquire two fundamentally different phase factors. One is of dynamical origin, it depends on the energy-dependent wave vector $k(E)$ of the electrons and on the length asymmetry between the two arms, whereas the other one is a geometrical phase arising from the presence of a perpendicular magnetic field to the sample, giving rise to a magnetic flux threading the ring $\Phi_{AB}$. The momentum $k(E)$ can be controlled by an external gate voltage $V_g$, whereas the magnetic flux $\Phi_{AB}$ is controlled by an external magnetic field $B$. The total transmission probability as a function of these parameters was derived in Ref.~[\onlinecite{Haack19}] and optimal parameters' ranges were estimated. In this work, we focus on two key parameters, the external gate voltage $V_g$ and the magnetic flux $\Phi_{AB}$, and fix the other parameters based on Ref.~[\onlinecite{Haack19}]. We recall below the energy-dependent expression of the AB ring transmission probability $T_{AB}(E)$, highlighting its dependence on $V_g$ and on the AB flux normalized by the flux quantum $\Phi_0 = h/e$, $\varphi \equiv 2 \pi \Phi_{AB}/\Phi_0$:
\bb
\label{eq:T_AB}
T_{AB}(E, V_g, \varphi) &=& \frac{ f_1(E, V_g)+ f_2(E, V_g)  \cos \varphi }{1  + \big[ f_3(E, V_g) -   f_4(E, V_g) \cos \varphi \big] ^2 }\,. \nonumber \\
&&
\ee
The exact expressions for the functions $f_i, i= 1, \ldots, 4$ are provided in App.~\ref{app:trans}. In Ref.~[\onlinecite{Haack19}], the functioning of this device was investigated in the linear response regime to a temperature bias across the AB ring, $\Delta T = T_h - T_c$. The linear response regime is valid when $\Delta T \ll (T_h  + T_c)/2$ and amounts to neglect higher-order non-linear terms in $\Delta T$. The authors showed that quantum transport through the device can be fully tuned by controlling the AB flux $\varphi$. Optimal working regime for thermoelectricity in the linear response regime could be achieved with low values of $\epsilon$, corresponding to the T-junctions behaving as two resonant tunnel barriers, similar to the Breit-Wigner regime in a Fabry-P\'erot interferometer. In this following work, we extend our investigation of this fully tunable phase-coherent thermoelectric device to the non-linear response regime, beyond small temperature bias.

\begin{figure}[t]
 \centering 
\includegraphics[width=\columnwidth]{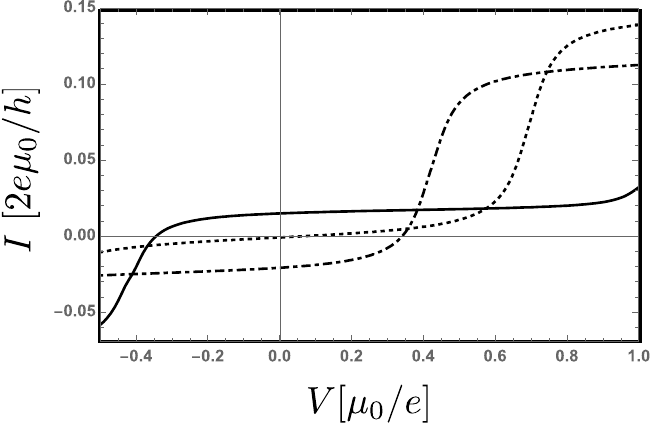}
\caption{$I-V$ characteristics in the non-linear regime for $T_h = 2$ K; $T_c=0.05$ K, $\varphi = 2 \pi \Phi_{AB}/\Phi_0 = \pi$, $\mu_0 = 10^{-22}$ J, $\epsilon=0.1$, and $\delta \tau = 0.3$. The curves correspond to $V_g= 0.63$ mV (solid), $V_g=1.1$ mV (dashed), and $V_g=1.5$ mV (dash-dotted). Thermovoltages for different $V_g$ correspond to the crossing points with the $x$-axis, \textit{i.e.} for $I=0$.}
\label{fig:2}
\end{figure}

\section{AB thermoelectricity in the non-linear regime}
\label{sec:open}

We start with the thermoelectric properties of the AB ring in an open-circuit configuration. Thermoelectricity refers to the ability of a device to exploit a voltage bias to induce a heat current (Peltier effect) or vice versa, to turn a temperature bias into an electrical current (Seebeck effect). Here, we focus on the latter, and will characterize the behavior of the Seebeck coefficient $S$ as a function of the AB ring's parameters. This coefficient tells us how much voltage the device is able to develop from a given temperature bias at zero charge current, $I=0$. Because this voltage is determined at zero charge current and originates from a temperature gradient, it was named \textit{thermovoltage}, and will be labeled $V_{th}$ in the following. In the linear response regime, the Seebeck coefficient $S_{l}$ is simply defined as the ratio between the thermovoltage $V_{th}$ and the small temperature difference $\Delta T$ at zero charge current:
\bb
S_{l}= \frac{V_{th}}{\Delta T}\Bigg\vert_{I=0}\,.
\ee
In general, this thermovoltage can also be built in presence of a voltage bias $V$ between the two contacts. In contrast, in the non-linear response regime, it is the differential Seebeck coefficient or differential thermopower $S_{nl}$ which characterizes the thermoelectric response to a temperature difference $\Delta T$ (not necessarily small in this regime).
\bb
\label{eq:S_nl}
S_{nl}= \frac{\partial V_{th}}{\partial \Delta T}\Bigg\vert_{I=0}\,.
\ee
This definition is in full analogy with the differential resistance in the non-linear response regime. In this work, we follow a scattering-matrix approach to compute the currents, in particular the charge current for estimating the thermo-voltage \cite{Lesovik11, Benenti17}:
\bb
\label{eq:current1}
I (V, V_g) &=& \frac{2 e}{h} \int_{-\infty}^{\infty} dE \, T_{AB}(E, U, V_g) \, \nonumber \\
&& \left[ n(E, \mu_L, T_h) - n(E, \mu_R- e V, T_c)\right]\,,
\ee
where $U$ describes the internal electrostatic potential of the AB ring and allows for taking into account interactions that may arise in the non-linear response regime. Whereas $U$ is taken at its equilibrium value $U_{eq}$ in the linear response regime, it may depend on external potentials applied onto the sample or to the reservoirs in general. The notation $n(E, \mu_i, T_i) = (1+ e^{(E-\mu_i)/(k_B T)})^{-1}$ corresponds to the Fermi distribution of hot (left) or cold (right) contact and $k_B$ the Boltzmann constant.

Equation~\eqref{eq:current1} is valid in both linear and non-linear response regimes. In this work, we investigate the non-linear response regime of the AB ring, assuming no interaction, hence no change in the internal electrostatic potential, \textit{i.e.} we consider it at its equilibrium value $U_{eq}$ \cite{Buttiker05}. Within this work, we justify this assumption as follows: i) no confined region such as a quantum dot is considered in our model, such that we avoid the Coulomb interaction  in this case \cite{Yacobi95, Hackenbroich96, Blanter97}, and ii) possible small modifications in $U$ can probably be compensated through the external gate voltage $V_g$ that we consider. We propose in this work to not consider explicitely the effect of interactions, hence to assume  $T_{AB}(E, V_g)$ as given by Eq.~\eqref{eq:T_AB} and $\mu_L = \mu_0$ and $\mu_R = \mu_0 + e V_{th}$ in the steady-state regime. Both contacts are initially set to the same chemical potential $\mu_0$ and we have to account for $V_{th}$ in the steady state.  A gauge invariant form of Eq.\eqref{eq:current1} is then
\bb
\label{eq:current2}
I (V, V_g) &=& \frac{2 e}{h} \int_{-\infty}^{\infty} dE \, T_{AB}(E, V_g) \, \nonumber \\
&& \left[ n(E, \mu_0, T_h) - n(E, \mu_0- e V, T_c)\right]\,,
\ee
A full derivation of higher order correction terms for the thermoelectric response is left for future investigation and could be conducted following the works \onlinecite{Lopez03, Hernandez09}.

\begin{figure*}[t]
	\centering 
	\includegraphics[width=0.9\linewidth]{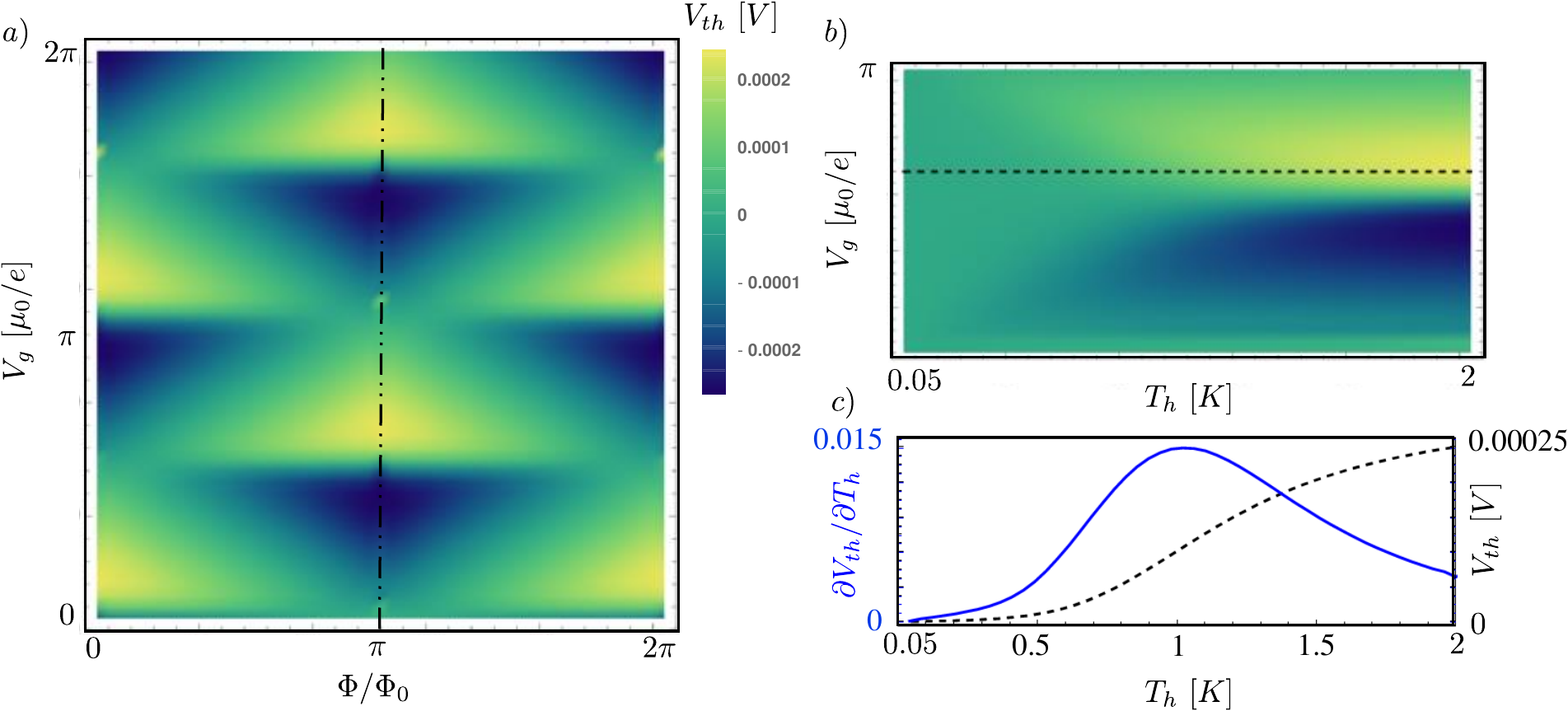}
	\caption{ Thermoelectric characteristics of the AB ring in the open-circuit configuration. \textbf{Panel a)}. Color plot of the thermovoltage $V_{th}$ developed across the AB interferometer vs magnetic flux $\varphi/2\pi$ and gate voltage $V_g$ calculated for $T_{c}=0.05$ K, $T_{h}=2$K, $\epsilon = 0.1, \delta \tau =0.3$. Numerical result with steps of $2\pi/50$ for both parameters. \textbf{Panel b)}. Color plot of thermovoltage $V_{th}$ as a function of the hot temperature $T_h$ and $V_g$, calculated at $\varphi=\pi$ and for fixed $T_{c}=0.05$ K. Dashed line indicates a cut at $V_g = 1.26$ mV. \textbf{Panel c)}. Differential thermopower (Seebeck coefficient in the non-linear regime), shown in blue solid curve with scales on the left vertical axis. It is calculated along the thermovoltage cut indicated as a dashed black curve in Panels b and c, left vertical axis.   
}
\label{fig:3}
\end{figure*}

Let us now return to the thermovoltage developed through this AB ring, in response to a temperature bias. By definition, $V_{th}$ is the solution of
\bb
\label{eq:Vth}
I (V_{th}, V_g) = 0\,,
\ee
with the current gievn by Eq.~\eqref{eq:current2}. The thermovoltage $V_{th}$ corresponds to the crossing points of the $I-V$ curves with the x-axis in Fig.~\ref{fig:2}. Here, the $I-V$ curve shows the steady-state charge current in response to a temperature bias as a function of the voltage $V$ for three different gate voltages $V_g$ (solid, dashed and dashed-dotted curves). We have fixed the magnetic flux $\varphi = \pi$, as well as the temperature bias across the sample, $T_h = 2$ K, $T_c = 0.05$ K. It can already be seen that the external voltage $V_g$ provides an important tunability for controlling $V_{th}$. To further explore the thermoelectric response of the AB ring in the non-linear regime, we show in Fig.~\ref{fig:3} the thermovoltage as a function of the gate voltage and magnetic flux (panel a), as a function of the temperature bias (panel b as a function of $T_{h}$, $T_{c}$ being fixed). The main features to retain from these figures is the high tunability of the thermovoltage with all parameters of the AB ring, and the possibility to switch on and off the thermoelectric response of the AB ring with the magnetic flux (see panel a).

In Fig.~\ref{fig:3} panel c, we show  the differential Seebeck coefficient, see Eq.~\eqref{eq:S_nl}. Remarkably, the differential Seebeck coefficient can be two orders of magnitude larger than the one in the linear response regime: the maximum of the blue curve reaches 15 mV/K, whereas the maximum attainable in the linear response regime was of the order of 300 $\mu$V/K \cite{Haack19}. This confirms the excellent thermoelectric response of this phase-coherent mesoscopic device, and motivates the rest of this contribution where we investigate its behavior as a quantum heat engine in a closed-circuit configuration.


\begin{figure*}[t]
	\includegraphics[width=0.9\linewidth]{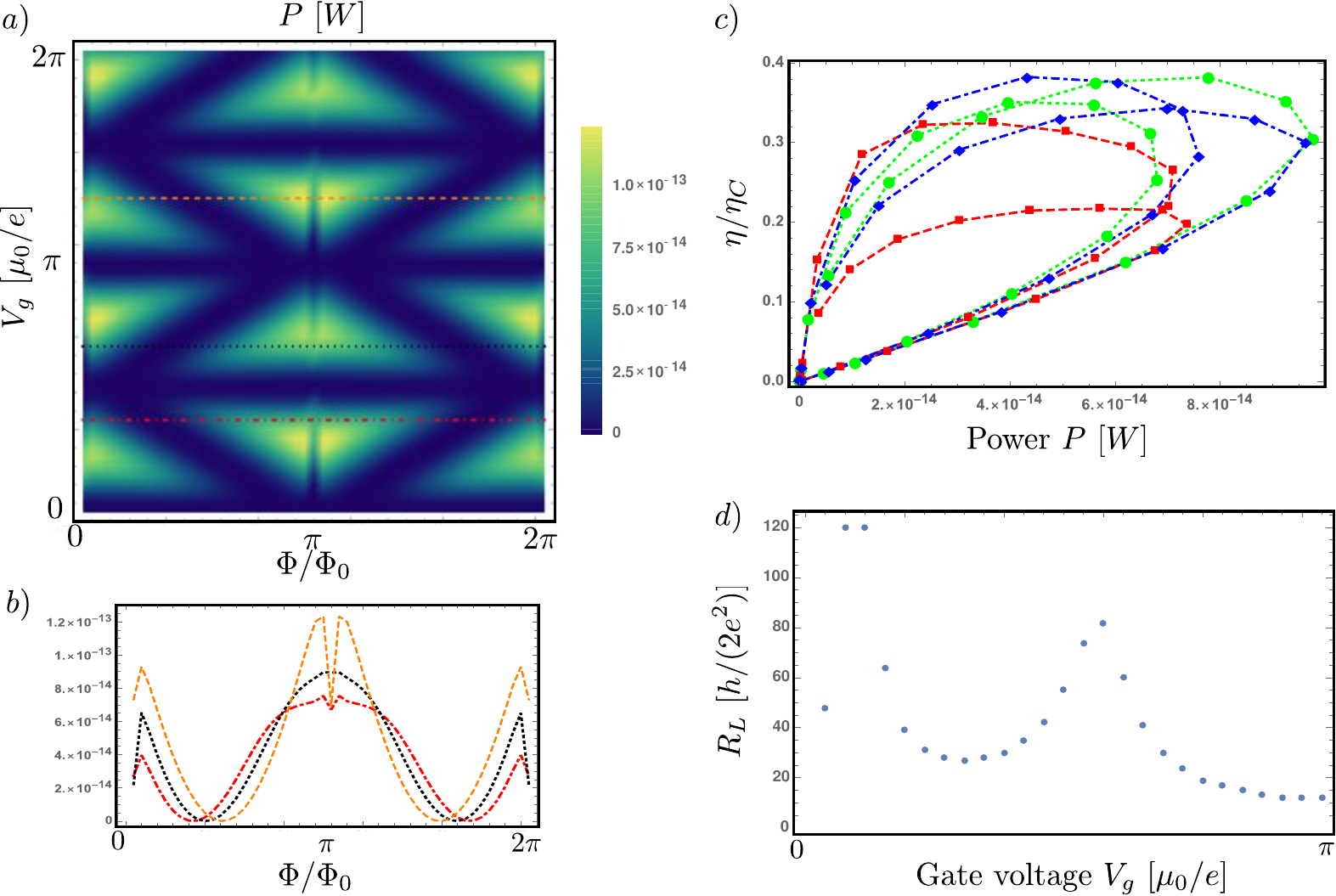} 
	\caption{Closed-circuit characteristics of the AB heat engine. \textbf{Panel a)}. Color plot of the power $P = (V_{th}^{(cl)})^2 / R_L$ as a function of gate voltage $V_g$ and magnetic flux $\varphi$. \textbf {Panel b)} Cuts of power as a function of magnetic flux for different values of the gate voltage, $V_g = 0.79$ mV (red dot-dashed), $V_g = 1.4 $ mV (black dotted) and $V_g = 2.7$ mV (orange dashed). Cuts are also shown on panel a). Non-linear behaviour of power Values for the other parameters for panels a) and b): $T_h = 2$ K, $T_c = 0.05$ K, $\epsilon = 0.1$, $\delta \tau = 0.3, R_L = 20 \, R_q$. \textbf{Panels c)}. Parametric plot of the efficiency normalized by the Carnot efficiency $\eta_C$ versus power $P$, as a function of the gate voltage $V_g = [0, \pi (\mu_0/e)]$ in steps of $\Delta V_g = 2 \pi / 50 (\mu_0/e)$. Different colors correspond to different load resistances: green (circles) for $R_L = 15 R_q$, blue (diamonds) for $R_L = 20 R_q$ and red (squares) for $R_L = 40 R_q$. Values of the other parameters are the same as for panels a) and b) with fixed magnetic flux, $\varphi = \pi$. \textbf{Panel d)}. Optimal load resistance as a function of $V_g$, for maximizing efficiency and power, see main text.
	}
	\label{fig:4}
\end{figure*}

\section{AB thermoelectricity in a closed-circuit configuration}
\label{sec:closed}

The closed-circuit configuration requires additional careful analysis for assessing correctly the thermovoltage and heat and charge currents. When the circuit is closed with a load resistance as shown in Fig.~\ref{fig:model} panel b), the Kirchhoff's laws have to be satisfied, ensuring charge and energy conservation: the sum of the charge currents flowing through the two branches of the circuit, one with the AB ring, the other one with the load resistance $R_L$, has to be equal to 0. As a consequence, the thermovoltage in this new configuration is obtained by solving a different equation as compared to Eqs.~\eqref{eq:current2} and \eqref{eq:Vth}, which reads:
\bb
&&\frac{2 e}{h} \int_{-\infty}^{\infty} dE T_{AB}(E) \left[ n(E, T_h) - n(E+ e V_{th}^{(cl)}, T_c) \right] = - \frac{V_{th}^{(cl)}}{R_L}\,, \nonumber \\
&&
\ee
and its solution for the thermovoltage is denoted $V_{th}^{(cl)}$ for clarity. We compute numerically values for the thermovoltages in the closed-circuit configuration as a function of gate voltage $V_g$ and magnetic flux $\varphi$, for fixed temperatures, fixed imbalance and fixed transmission probabilities of the T-junctions. This allows us to compute the heat current flowing through the AB ring in the non-linear response regime according to Refs.~\onlinecite{Houten92, Lesovik11, Lopez16}. Let us precise that in the context of the AB ring operated as heat engine, we have to consider the heat current flowing into the hot contact, $J_h$, as it corresponds to the resource for the engine\cite{Benenti17}. It corresponds to the energy flow that electrons can dissipate into the hot reservoir, taking as reference the chemical potential of that same reservoir, here $\mu_0$:
\bb
\label{eq:heat}
J_h &=& \frac{2}{h} \int_{-\infty}^{\infty} dE (E - \mu_0)\, T_{AB}(E) \left[ n(E, T_h) - n(E+ e V_{th}^{(cl)}, T_c)\right]\,, \nonumber \\
&&
\ee
For clarity, we recall that heat current is not conserved, $J_h \neq J_c$ with $J_c$ the heat current flowing into the cold contact. In the case of the AB ring operated as heat engine with the building up of a thermovoltage on the right (cold) lead, conservation law instead reads:
\bb
J_h + J_c = V_{th}^{(cl)} I\,,
\ee
with
\bb
J_c &=& \frac{2}{h} \int_{-\infty}^{\infty} dE (E - (\mu_0 - e V_{th}^{(cl)})\, \nonumber \\
&& T_{AB}(E)  \left[ n(E+ e V_{th}^{(cl)}, T_c) - n(E, T_h) \right]\,.
\ee
We also compute the power $P = (V_{th}^{(cl)})^2 / R_L$ generated by the AB heat engine, which allows us in the next section to discuss several figures of merit for the AB heat engine in the non-linear response regime. The load resistance is typically expressed as a multiple of the quantum of resistance $R_q = h/(2e^2)$. The numerical integration over energy for computing the heat current was performed over ten times the Fermi energy $\mu_0$. Let us remark that it takes values of the same order of magnitude as the one in the linear response regime for $T = 1$ K, namely of the order of 0.1 -1 pW, see App.~\ref{app:heat}. Figure~\ref{fig:4} panel a) shows the power for a fixed value of the load resistance, $R_L = 20 \, R_q$ as a function of gate voltage and magnetic flux, and panel b) highlights the power characteristics for different gate voltages as a function of the magnetic flux. The cuts evidence the possibility to fully turn on and off the heat engine, from 0 to maximal power.


\section{Power and efficiency of the AB heat engine in the non-linear response regime}
\label{sec:eff}

We now investigate the power versus efficiency trade-off in the non-linear regime, as a function of the gate voltage $V_g$ and the load resistance $R_L$. Given a temperature bias, we determine the optimal load resistance to maximize both power and efficiency, the latter being simply defined as the ratio of power (output of the engine in $[W]$) and heat current into the hot reservoir (input of the engine, also in $[W]$):
\bb
\eta = P/J_h\,.
\ee
In Fig.~\ref{fig:4} c), we show the Lasso-type parametric plot of efficiency versus power, that evidences the trade-off between both quantities. A heat engine with maximal efficiency, Carnot efficiency, does not produce any power as it corresponds to the point of reversibility for the machine \cite{Benenti17}. Hence a compromise must be found or determined depending on the primary goal of the engine (to produce more power with less efficiency or be more efficient with less output power). Here, we show that the AB heat engine can be easily tuned through external gate voltage $V_g$ (panel c) to optimize either efficiency or power. These two quantities also depend on the load resistance, as evidenced by the different colored curves in the same panel. For the AB heat engine, it seems that a load resistance of $R_L \sim 15 - 20\, R_q$ may be optimal given the values we have fixed for the other parameters. Interestingly, an efficiency up to $\sim$ 40$\%$ of Carnot efficiency can be achieved by adjusting $V_g$ and $R_L$. This value has to be compared to the efficiency in the linear response regime for similar temparture gardient, where the authors showed a maximal efficiency of the order of $30\%$ the Carnot efficiency \cite{Haack19}. The increase in efficiency can directly be attributed to the non-linear response of the AB engine, as similar temperature bias accross the setup were considered (while keeping the same parameters for asymmetry, T-junctions parameter and magnetic flux). In panel d), we show the optimal load resistance to maximize both efficiency and power, as a function of the gate voltage. To achieve this, we maximize $P + \eta$,over $R_L$ for fixed $V_g$ and fixed temperature gradients. This procedure does not reflect the absolute optimal $R_L$ as a function of all parameters (of interest in possible future experiments), but rather the high tunability of the AB ring as heat engine to achieve sizeable power and efficiency in the quantum regime.

\section{Conclusions and  perspectives}

In conclusion, we have analyzed a coherent mesoscopic heat engine consisting of an Aharonov-Bohm quantum interferometer operated in the full non-linear regime. The device thermoelectric response turns out to be sizable, and the interferometer is a able to provide a thermopower which is about 50 times larger than in the same structure operated in the linear regime. Moreover,  also the heat engine thermoelectric efficiency at maximum power is somewhat large, obtaining values as high as $\sim 40\%$   of the Carnot efficiency in the linear regime. The AB heat engine can provide magnetic flux- and electrostatic-driven control of charge and heat current as well as of its thermoelectric response under physical conditions which are accessible from the experimental point of view.
Suitable candidates for the implementation of the Aharonov-Bohm quantum heat engine are represented by metallic or GaAs/AlGaAs two dimensional electron gas heterostructures \cite{Webb85, Washburn86, Yacobi96, Yamamoto12}, which are expected to yield devices with robust performances over a wide range of system parameters. Future experiments will also stimulate further theoretical work to address the effect of interactions that may induce corrections due to a modified internal electrostatic potential. The above results show that an Aharonov-Bohm loop operated as a quantum heat engine in the non-linear regime represents a prototypical platform for the implementation of a unique class of phase-tunable thermoelectric quantum machines operating at cryogenic temperatures. Yet, in the context of quantum technologies \cite{Acin18}, this coherent structure might be at the core of a number of innovative thermolectric quantum devices, for instance, highly-sensitive photon sensors where radiation-induced heating one of the electrodes forming the system is detected via the resulting thermovoltage.

\begin{acknowledgments}
We acknowledge stimulating discussions with Rosa Lopez about the non-linear response regime within a scattering-martix approach. GH acknowledges support from the Swiss National Science Foundation, through the PRIMA grant PR00P2$\_$179748 and the National Centers for Competences and Research for Quantum Science and Technology (QSIT) and SwissMap. FG acknowledges the European Research Council under Grant Agreement No. 899315-TERASEC, and  the  EU’s  Horizon 2020 research and innovation program under Grant Agreement No. 800923 (SUPERTED) and No. 964398 (SUPERGATE) for partial financial support.

\end{acknowledgments}

\section*{Author Declarations} The authors have no conflicts to disclose.

\section*{Data Availability Statement}

The techniques and values used to generate the data that support the findings of this study are available within the article and from the corresponding author upon reasonable request.

\appendix

\section{Transmission probability of the AB ring}
\label{app:trans}

In this section, we provide the expressions of the functions $f_i$ in Eq.~\eqref{eq:T_AB} of the transmission probability of the AB ring. These coefficients are determined and expressed according to Ref.~[\onlinecite{Haack19}], where the authors previously derived $T_{AB}$ in full generality:
\bb
 f_1(E, V_g) &=& 1- \cos\chi \cos \delta \chi \\
 f_2(E, V_g) &=& \cos \delta \chi  - \cos \chi \\
f_3(E, V_g) &=& \frac{2(1-\epsilon) \cos \chi  - (1-\epsilon - \sqrt{1-2 \epsilon})\cos \delta \chi }{2 \epsilon \sin \chi} \nonumber \\
&& \\
f_4(E, V_g) &=& \frac{ \big(1-\epsilon + \sqrt{1-2 \epsilon}\big) }{2 \epsilon \sin \chi}\,,
\ee
with
$\chi = \chi_1 + \chi_2$ and $\delta \chi = \chi_1 - \chi_2$ and $\chi_i = k_i L_i$ are the dynamical phases that electrons acquire while traveling in each arm $i=1,2$. Here $k_i$ is the energy-dependent wave vector and $L_i$ the length of the corresponding arm
\bb
k_1(E) &=& \tilde{k}_{\mu_0} + \frac{E-{\mu_0}}{\hbar v_d}\, , \,  k_2(E) =  \tilde{k}_{\mu_0} + \frac{E-(\mu_0+eV_g)}{\hbar v_d}\,.
\ee 
For an asymmetric AB ring, the difference and sum of the dynamical phases then read:
\bb
\label{eq:chi}
\!\!\!\!\chi_1 + \chi_2 &=& (2 L + \delta L) \left(\tilde{k}_{\mu_0} + \frac{E-\mu_0}{\hbar v_d} \right)  - \frac{e V_g L}{\hbar v_d} \label{eq:chis}\\
\!\!\!\!\chi_1 - \chi_2 &=& \delta L \left(\tilde{k}_{\mu_0} + \frac{E-\mu_0}{\hbar v_d} \right)  + \frac{e V_g L}{\hbar v_d}\,. \label{eq:chid}
\ee
Here we have defined $L_1 \equiv L + \delta L$, $L_2 \equiv L$, $v_d$ is the electronic drift velocity and the Fermi wave vector is tuned through energy offsets applied onto each arm ($k_{\mu_0} \rightarrow \tilde{k}_{\mu_0}$).

\section{Heat current}
\label{app:heat}

\begin{figure}[t]
\centering 
\includegraphics[width=1\linewidth]{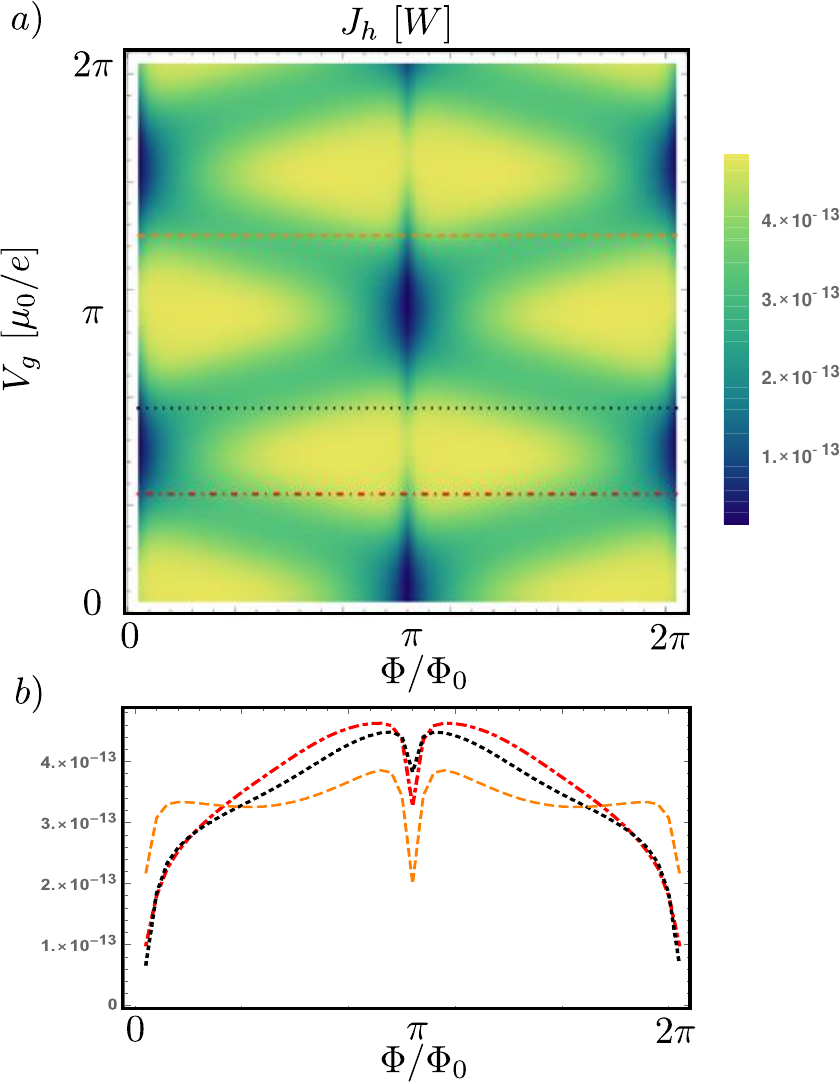}
\caption{Heat current through the AB heat engine in a closed-circuit configuration. \textbf{Panel a)}. Color plot of the heat current $J_h$ (see Eq.~\eqref{eq:heat}), as a function of gate voltage $V_g$ and magnetic flux $\varphi$.  
\textbf{Panel b)}. Cuts of heat current as a function of magnetic flux for different values of the gate voltage, $V_g = 0.79$ mV (red dot-dashed), $V_g = 1.4 $ mV (black dotted) and $V_g = 2.7$ mV (orange dashed). Other parameters are fixed: $T_h = 2$ K, $T_c = 0.05$ K, $\epsilon = 0.1$, $\delta \tau = 0.3$
}\label{fig:app_heat}
\end{figure}

For completeness, we provide with Fig.~\ref{fig:app_heat} a color plot of the heat current as a function of voltage gate and magnetic flux, similar to the power shown in Fig.~\ref{fig:4} panels a and b. We also indicate the same cuts for values of $V_g$ given in the main text.


\end{document}